\documentclass[a4paper,11pt]{article}
\usepackage[margin=2.5cm]{geometry}
\usepackage{graphicx}
\usepackage{epsfig}
\usepackage{amsmath}
\usepackage{amssymb}
\usepackage{cite}
\usepackage{float}
\usepackage[font={footnotesize,it}]{caption}
\usepackage{authblk}
\usepackage[utf8]{inputenc}
\usepackage[titletoc,toc,title]{appendix}
\usepackage{epstopdf}
\begin{document}
\begin{titlepage}

\title{Entanglement thermodynamics for charged black holes}

\author[1]{Pankaj Chaturvedi \thanks{\noindent E-mail:~  cpankaj@iitk.ac.in}}
\author[2]{Vinay Malvimat\thanks{\noindent E-mail:~ vinaymm@iitk.ac.in }}
\author[3]{Gautam Sengupta\thanks{\noindent E-mail:~  sengupta@iitk.ac.in}}

\affil[1,2,3]{
Department of Physics,

Indian Institute of Technology Kanpur,

Kanpur 208016, INDIA}

\maketitle

\abstract{
\noindent

The holographic quantum entanglement entropy for an infinite strip region of the boundary for the field theory dual to charged black holes in ${\cal A}dS_{3+1}$ is investigated. In this framework we elucidate the low and high temperature behavior of the entanglement entropy pertaining to various limits of the black hole charge. In the low temperature regime we establish a first law of entanglement thermodynamics for the boundary field theory. }

\end{titlepage}
\tableofcontents
\pagebreak
\section{Introduction}
Quantum correlations play an important role in studying various aspects of many-body physics pertaining to condensed matter systems, statistical mechanics, quantum information and quantum gravity. In this regard, entanglement entropy is a crucial quantity which provides a measure for quantum correlations in a bipartite quantum system.  Such a system is described as the union of a subsystem $A$ that is being studied and which is correlated with the rest of the system described by its complement $A_c$. For this system the full Hilbert space ${\cal H}$ may  be expressed as a tensor product of the Hilbert spaces of  $A$  and  $A_c$ as  ${\cal H}={\cal H}_{A}\otimes {\cal H}_{A_c}$. At zero temperature, the whole system can be represented by a pure state $|\psi\rangle~\epsilon~{\cal H}$, with the density matrix given by $\rho=|\psi\rangle\langle\psi|$. However, if the subsystems $A$ and $A_c$ are entangled then the full state $|\psi\rangle$ cannot be expressed as a tensor product of any two states of the subsystems i.e  $|\psi\rangle\neq |\phi\rangle_{A}$ $\otimes$ $|\lambda\rangle_{A_c}$ where, $|\phi\rangle_{A}$ and $|\lambda\rangle_{A_c}$ are states in the bases of $A$ and $A_c$ respectively. This naturally also implies that the density matrix $(\rho)$ of the full system cannot be expressed as the tensor product of $\rho_A$ and $\rho_{A_c}$ i.e. $\rho\neq\rho_{A}\otimes\rho_{A_c}$.  The reduced density matrix $(\rho_A)$ for the subsystem $A$ is obtained via tracing $\rho$ over the degrees of freedom of $A_c$ giving $\rho_{A}=Tr_{A_c}\rho$ and the entanglement entropy of the subsystem $A$ is defined as the corresponding Von Neumann entropy, $S_{A}=-Tr_{A}(\rho_{A}\log\rho_{A})$. Entanglement entropy is know to have several important properties such as the strong subadditivity condition for two different regions $A_1$ and $A_2$ of a quantum system   ($S_{A_1}+S_{A_2}\geq S_{A_1 \cup A_2} + S_{A_1\cap A_2}$) \cite{araki1970}.

  To obtain the entanglement entropy for  1+1 dimensional  conformal field theories ($CFT_{1+1}$), a method was developed by Calabrese and Cardy using what is known as the replica trick. They showed that the entanglement entropy for a $CFT_{1+1}$ exhibits an universal logarithmic behavior \cite{Calabrese:2004eu,Calabrese:2009qy,Holzhey:1994we,Callan:1994py,Vidal:2002rm,Latorre:2003kg}. The forementioned authors further developed techniques to study quantum entanglement of $CFT_{1+1}$ in various static as well as dynamic scenarios \cite{Sotiriadis:2010si,Cardy:2011zz,Calabrese:2012ew,Calabrese:2014yza,Cardy:2015xaa}. However, very little is known about CFTs in higher dimensions, where, only the case of quasi free fermions and bosons has been studied rigorously in\cite{Cramer:2006mje,PhysRevA.73.012309,PhysRevLett.94.060503,PhysRevLett.98.220603,PhysRevLett.96.100503,PhysRevLett.96.010404}.
  
Further insight into the entanglement entropy for quantum field theories was facilitated through the $AdS/CFT$ correspondence which relates  $(d)$-dimensional strongly coupled boundary quantum field theory at its conformal fixed point in a large N limit, to a theory of weakly coupled gravity in $(d+1)$-dimensional bulk space-time \cite{Maldacena:1997re,Gubser:1998bc,Witten:1998qj,Witten:1998zw,Aharony:1999ti}. Using this correspondence, Ryu and Takyanagi proposed a conjecture to holographically evaluate the entanglement entropy of a $(d)$-dimensional quantum field theory \cite{Ryu:2006bv,Ryu:2006ef} which was later generalized for time dependent backgrounds by Hubeny-Rangamani-Takayanagi (HRT) in \cite{Hubeny:2007xt}. This holographic prescription for obtaining the entanglement entropy $S_A$ for a region $A$ (enclosed by the boundary $\partial A$) in the $(d)$-dimensional boundary field theory involves the computation of the area of the extremal surface (denoted by $\gamma_A$) extending from the boundary $\partial A$ of the region $A$ into the $(d+1)$-dimensional bulk such that $S_A$ is given by
\begin{equation}
 S_A= \frac{Area(\gamma_A)}{4G_N^{d+1}},\label{EEarea}
\end{equation}
 where $G_N^{(d+1)}$ is the gravitational constant of the bulk \cite{Nishioka:2009un}. Besides this, they correctly reproduced the logarithmic behavior of the entanglement entropy for a $(1+1)$-dimensional boundary $CFT$ in $AdS_{2+1}/CFT_{1+1}$ and also confirmed the validity of the strong subadditivity inequalities in holographic CFTs \cite{Nishioka:2009un,Headrick:2007km,Callan:2012ip,Wall:2012uf,Caceres:2013dma,Chen:2014hua,Lashkari:2014kda}. One of the several applications of the conjecture involves the study of the entanglement entropy for a boundary $CFT$ at finite temperature which is dual to a AdS-Schwarzschild black hole in the bulk \cite{Cadoni:2009tk,e12112244,Hubeny:2012ry,Fischler:2012ca}. The results of which were subsequently used to study the low and the high temperature behavior of the entanglement entropy. This in turn led to the proposal of an analogous {\it ``First law of thermodynamics''} like relation for the entanglement entropy by the authors in \cite{Bhattacharya:2012mi, Allahbakhshi:2013rda}. 

Another class of boundary field theories that have been studied using the Ryu and Takyanagi conjecture are those which are dual to charged AdS black holes in the bulk \cite{Tonni:2010pv,Hubeny:2012ry,Caputa:2013eka,Mansoori:2015sit,Caputa:2013lfa}. In particular, the authors in \cite{Tonni:2010pv} have studied the entanglement entropy for the boundary field theory dual to a Reissner-Nordstrom-AdS black hole in $(d+1)$-dimensional $AdS$ spacetime. For such a boundary field theory at finite temperature and charge density, the issue of entanglement thermodynamics was also addressed in \cite{Blanco:2013joa,Park:2015afa,He:2014lfa}. However,  a complete description of the temperature and charge dependence of the entanglement entropy in this case remains to be studied. In this article using the framework of  \cite{Fischler:2012ca}, we comprehensively investigate the temperature and charge dependence of the entanglement entropy for a strip like region in the  boundary field theory which is dual to $RN$ black holes in $AdS_4/CFT_3$ setup. Furthermore, we also explore the issue of  entanglement thermodynamics pertaining to such strongly coupled boundary field theories.

This article is organized as follows. In section 1, we review the Ryu-Takyanagi conjucture and describe the setup for computing the holographic entanglement entropy for boundary field theory dual to planar uncharged black holes. In section 2, we state the holographic prescription for computing the entanglement entropy for boundary field theories dual to charged Reissner-Nordstrom black holes in a $AdS_4/CFT_3$ scenario. In section 3, we compute the entanglement entropy for boundary field theory dual to extremal and non-extremal charged black holes in the small charge regime and obtain the ``First law of entanglement thermodynamics" at low temperatures. In the same section, we also study the low and high temperature behavior of entanglement entropy in the small charge regime for the case of non-extremal black holes in the $AdS_4$ bulk spacetime. In section 4, we present the entanglement entropy for boundary field theory dual to extremal and non-extremal charged black holes in the large charge regime. Finally in section 5, we summarize out results and findings.
\section{Review of holographic entanglement entropy for uncharged planar black holes}
\begin{figure}[H]
\centering
\includegraphics[width =3in,height=2.2in]{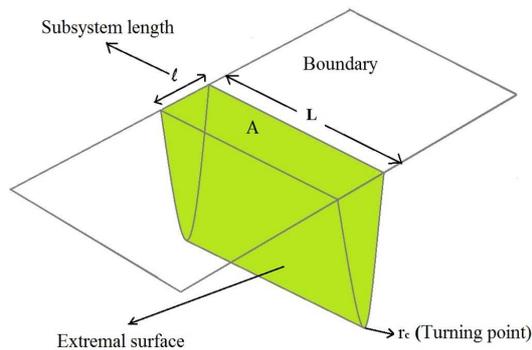}
\caption{\label{fig:subsystem}Schematic of extremal surface anchored on the subsystem that lives on the boundary}\end{figure}

 In this section, we review the method adopted in \cite{Fischler:2012ca} for the computation of finite temperature entanglement entropy for a boundary field theory dual to a AdS-Schwarzschild black hole in the $AdS_{3+1}$ bulk space time, and also describe its high and low temperature behavior. For our purpose, we will consider a $AdS_4/ CFT_3$ scenario with restriction to the Poincar\'{e} patch of $AdS_4$. The boundary conformal field theory considered is partitioned into a subsystem (A) which is geometrically a long strip and is entangled with the rest of the system as shown in the fig.(\ref{fig:subsystem}).

\pagebreak
In Poincar\'{e} coordinates, the metric for a black hole with a planar horizon in $AdS_{3+1}$ spacetime is given as

\begin{equation}
ds^2= -\frac{r^2}{R^2}f(r)dt^2+\frac{R^2dr^2}{r^2f(r)}+\frac{r^2}{R^2}d\vec{x}^2,\label{metric}
\end{equation}
 where $R$ is the $AdS$ length scale which is later set to 1  and the components of the vector $\vec{x}$ corresponds to $\{x,y\}$. In the boundary field theory we have chosen the subsystem $(A)$ as a long strip defined by $x\in\big[-\frac{l}{2},\frac{l}{2}]$, $y\in \big[\frac{L}{2},\frac{L}{2}]$ as shown in fig.(\ref{fig:subsystem}). The area of the surface anchored on the boundary of the subsystem $(A)$ may be then expressed as

\begin{equation}
{\cal A}=2 L\int^\infty_{r_{c}} \frac{ dr}{\sqrt{f(r)(1-\frac{r_{c}^{4}}{r^{4}}})}.\label{Earea1}
\end{equation}
here, $r_{c}$ is the constant of integration and represents the turning point of the extremal surface in the higher dimensional $AdS_4$ bulk spacetime. This quantity $r_c$ is obtained by inverting the equation of motion

\begin{equation}
  \frac{l}{2}=\int^\infty_{r_{c}} \frac{r_{c}^{2} dr}{r^{4}\sqrt{f(r)(1-\frac{r_{c}^{4}}{r^{4}})}}.\label{lrel}
\end{equation}

From eq.(\ref{Earea1}) it may be seen that the area integral is divergent as, $r\rightarrow \infty$ and has to be regularized by introducing an infrared cutoff ($r_b$) in the bulk $AdS$ spacetime. The holographic dictionary relates the UV cutoff of the boundary field theory($a$) to bulk IR cutoff  and these are inversely related through the $AdS$ length scale $R$ as, $r_b=\frac{R^2}{a}$. Furthermore, the finite part of entanglement entropy can then be used to study the high and low temperature behavior of entanglement entropy for the boundary field theory which is dual to Schwarzschild-AdS planar black hole as covered in \cite{Fischler:2012ca}  

\begin{equation}
 S_A^{finite}=S_A-S_A^{divergent}=\frac{{\cal A}^{finite}}{4G_{N}^{(d+1)}}.
\end{equation}

The integrals mentioned in equations (\ref{Earea1}) and (\ref{lrel}) have no known analytic solutions and several approximation  techniques have been used to study these. However, these methods were mostly valid for the low temperature regime. In
\cite{Fischler:2012ca}, the authors have used a novel expansion technique to compute the area integral eq.(\ref{Earea1}) order by order for both the low and high temperature regimes. In this article, we will also incorporate this method to obtain an analytic expression for the entanglement entropy of the boundary field theory dual to a bulk Reissner-Nordstrom-AdS black hole.

Note that the low temperature limit of the boundary field theory corresponds to a black hole with small horizon radius which implies the limit, $r_h\lll r_c$. In this limit the authors in \cite{Fischler:2012ca} showed that the leading contribution to the entanglement entropy $S_A$ arises from $S_A^{AdS}$ which is the entanglement entropy of the subsystem $(A)$ of the boundary field theory dual to the pure $AdS_{3+1}$ bulk. The explicit form of $S_A$ in the low temperature approximation can be written down as follows
 
\begin{equation}
 S_A=S_A^{AdS}+k(r_hl)^3.\label{lsasch}
\end{equation}

The low temperature behavior for Schwarzschild case was also studied in \cite{Bhattacharya:2012mi}, where the authors derive a first law like relation for the dual boundary field theory which is also known as ``First law of entanglement thermodynamics''. This law states that for a subsystem $A$ of the boundary field theory, the difference between the entanglement entropy of an excited state at a small non-zero temperature and the zero-temperature ground state ($\Delta S_A=S_A^{Temp\neq 0}-S_A^{Temp=0}$) is proportional to the change in the internal energy ($\Delta E_A$) of the subsystem $(A)$ as

\begin{equation}
\Delta S_A=\frac{1}{T_{ent}}\Delta E_A,
\end{equation}
The proportionality constant $T_{ent}$ is known as {\it entanglement temperature} and was shown to be inversely related to the subsystem length $(l)$ as ($T_{ent}=c \ l^{-1}$)
    
On the other hand  in \cite{Fischler:2012ca}, the high temperature behavior (corresponding to the limit, $r_h\rightarrow r_c$) of the entanglement entropy was shown to follow the form given below

\begin{equation}
S_A = c_0 Ll T^{2}+TL(c_1+c_2 \epsilon)+O[\epsilon^2],~~\epsilon\propto \exp(-k T l).\label{hsasch}
\end{equation}

The first term in the above expression for $S_A$ corresponds to the extensive thermal entropy of the subsystem $(A)$ as it scales with the area of the subsystem in the $(2+1)$-dimensional conformal boundary and as volume in higher dimensions. The $\epsilon$ corrections decrease exponentially with temperature and they scale as the length of the boundary of subsystem (A) and therefore correspond to entanglement between subsystem $(A)$ and the rest of the system. 
\section{Entanglement entropy of charged planar black holes}

In this section, using the Ryu-Takyanagi prescription we establish the framework for computing the entanglement entropy of a strip like region (referred to as subsystem $(A)$) in the  boundary field theory which is dual to $RN$ black holes in $AdS_4/CFT_3$ setup \footnote{Note that even though we restrict ourselves in the present work to  $AdS_4/CFT_3$, the techniques we use for computation are quite general. Therefore, we expect the generalization to higher dimensions to be straightforward.}. The metric of Reissner Nordstrom black hole in $AdS_4$ spacetime with a planar horizon is given as

\begin{eqnarray}
ds^2 &=& -\frac{r^2}{R^2}f(r)dt^2+\frac{R^2}{r^2f(r)} dr^2+\frac{r^2}{R^2}(dx^2+dy^2),\label{RNmetric}\\
f(r)&=& 1-\frac{M}{r^3}+\frac{Q^2}{r^4},\label{RNlapse}
\end{eqnarray}
where `R' is the AdS length scale which we set to 1 (R=1). The Hawking-temperature of planar $RN$ black hole in $AdS_4$ may be expressed as

\begin{equation}
T=\left.\frac{f'(r)}{4\pi}\right|_{r=r_h}=\frac{3 r_{h}}{4\pi}(1-\frac{Q^{2}}{3 r_{h}^{4}}).\label{RNtemp}
\end{equation}
 
Now using equations (\ref{lrel}) and (\ref{Earea1}) with the lapse function $(f(r))$ and metric given by eq. (\ref{RNmetric}) and eq.(\ref{RNlapse}) and the length and the area integral for the subsystem $(A)$ of the boundary field theory dual to $RN$ black hole in $AdS_4$ spacetime may be expressed as

\begin{eqnarray}
 \frac{l}{2}&=&\int^\infty_{r_{c}}\frac{ r_{c}^{2} dr}{r^4\sqrt{(1-\frac{r_{c}^4}{r^4})}}\Big(1-\frac{M}{r^3}+\frac{Q^2}{r^4}\Big)^{-\frac{1}{2}},\label{RNlrel}\\
 {\cal A}&=&2L \int^\infty_{r_{c}}\frac{  dr}{\sqrt{(1-\frac{r_{c}^4}{r^4})}}\Big(1-\frac{M}{r^3}+\frac{Q^2}{r^4}\Big)^{-\frac{1}{2}}.\label{RNEarea}
\end{eqnarray}

The equations (\ref{RNlrel}) and (\ref{RNEarea}) may then be used to determine analytically the entanglement entropy of the subsystem $(A)$ of the $(2+1)$-dimensional boundary field theory. It may be seen from eq.(\ref{RNlrel}) the subsystem length $l$ can be obtained as a function of $r_c$. We then invert the relation between $l$ and $r_c$ to obtain $r_{c}$ as a function of subsystem length $l$ and substitute it in expression for the area of the extremal surface given by the eq.(\ref{RNEarea}). This procedure determines the form of area of the extremal surface solely in terms of the subsystem length $l$, the black hole charge $Q$ and the black hole mass $M$. However, it is to be noted that in the bulk theory there are only two parameters namely the charge $Q$ and the mass $M$ of the black hole which are related to each other by the radius of horizon $(r_h)$ of the black hole as follows

\begin{equation}
 f(r_{h})=0 \Rightarrow M=\frac{r_{h}^4+Q^2}{r_{h}}.\label{MQrel}
\end{equation}

The condition $f(r_h)=0$, implies that the lapse function vanishes at the horizon $(r=r_h)$. Then using the relation in eq.(\ref{MQrel}) we re-express the lapse function $(f(r))$ in terms of the radius of horizon $r_h$ and the charge $Q$ of the black hole as follows

\begin{equation}
 f(r)=1-\frac{r_h^3}{r^3}-\frac{Q^2}{r^3r_h}+\frac{Q^2}{r^4}.\label{RNlapse1}
\end{equation}

Thus with this reparametrization of the lapse function $f(r)$ it can be said that the bulk theory is now characterized effectively by the charge $Q$ and the radius of horizon $r_h$. Next, in order to evaluate the integrals we make a change of variable from $r$ to $u=\frac{r_{c}}{r}$ in expressions of subsystem length $l$ and area ${\cal A}$ given by eq.(\ref{RNlrel}) and eq.(\ref{RNEarea}) which give us the following modified forms of the integrals

\begin{eqnarray}
 l&=&\frac{2}{r_{c}}\int _0^1\frac{u^2 \left(1-\frac{{r_h}^3 u^3}{{r_{c}}^3}-\frac{Q^2 u^3}{{r_{c}}^3 {r_h}}+\frac{Q^2 u^4}{{r_{c}}^4}\right)^{-\frac{1}{2}}}{\sqrt{1-u^4}}du, \label{eq4}\\
{\cal A}&=&2Lr_{c}\int _0^1\frac{\left(1-\frac{{r_h}^3 u^3}{{r_{c}}^3}-\frac{Q^2 u^3}{{r_{c}}^3 {r_h}}+\frac{Q^2 u^4}{{r_{c}}^4}\right)^{-\frac{1}{2}} }{u^2\sqrt{1-u^4}}du.\label{eq5}
\end{eqnarray}

The expression for subsystem length $l$ and extremal area ${\cal A}$ given by eq.(\ref{eq4}) and eq.(\ref{eq5}) can then be used to compute the entanglement entropy for the subsystem $A$ and study its behavior with temperature and charge of the $RN$ black hole which we will address in later sections. Furthermore note that for a boundary field theory dual to a charged black hole in the bulk AdS$_4$ spacetime the state space depends on two parameters namely the temperature $T$ (which is related to radius of horizon $r_h$) and the charge $Q$ of the black hole. Hence it is necessary to consider the bulk charged RN-AdS black hole in a specific ensemble. Here we choose to work in the canonical ensemble for which the charge $Q$ of the black hole remains fixed. Unlike the case of boundary field theory dual to Schwarzschild black hole in the $AdS_4$ bulk where the low and high temperature behavior of the entanglement entropy of subsystem $A$ is controlled only by the temperature $(T)$ of the black hole for the case of the boundary field theory dual to charged black hole  it is also dependent on the charge $Q$. This may be seen from the  extremality  condition for the $RN$ black holes in the bulk $AdS_4$ spacetime obtained from eq.(\ref{RNtemp}) as

\begin{equation}
r_{h}\geq\frac{\sqrt{Q}}{{3^{\frac{1}{4}}}}.\label{eq2}
\end{equation}

Here the equality is satisfied when the black hole is at zero temperature (i.e extremal) and the inequality refers to a non zero temperature for the charged AdS black hole. Therefore the second factor in the expression for the temperature given by eq.(\ref{RNtemp}) decides whether the black hole is close to extremality or not ( alternatively at low or high temperatures). Hence this factor appears recurrently in our calculations for the finite temperature entanglement entropy as will be shown in the later sections. It is also to be noted that for the case of boundary field theory dual to charged black holes the zero temperature ground state is dual to extremal black holes. Thus in order to obtain a ``first law of entanglement thermodynamics" we also investigate the entanglement entropy of the boundary field theory dual to extremal charged AdS black holes.

From the inequality in eq.(\ref{eq2}), it may be observed that the horizon radius $(r_h)$ is bounded from below by a quantity which proportional to the charge $Q$ of the black hole. Thus the range of possible values for the horizon radius is decided by the value of the charge $Q$.  More precisely, if the charge of the black hole is small then the radius of horizon $(r_h)$ may assume both small (low temperature regime) and large values (high temperature regime). However, if the charge of the black hole is large then $r_h$ can only assume large values (only the high temperature regime can be explored). In contrast to the case of non-extremal black holes, the radius of horizon $(r_h)$ for extremal black holes is directly related to the charge $Q$ of the black hole which may be seen from the equality condition in eq.(\ref{eq2}). Thus in the case of extremal black holes which are at zero temperature small or large charge will imply small or large radius of horizon $(r_h)$ respectively. So, in the light of facts mentioned above we will explore the behavior of entanglement entropy in distinct regimes of the value of the charge of the planar $RN$ black hole in the bulk$AdS_4$ spacetime. To elucidate this further  we will first explore the entanglement entropy of the subsystem $A$ of the boundary field theory dual to extremal and the non-extremal charged black holes in the small charge regime and subsequently continue our investigation to the large charge regime.


\section{Small charge regime}

In this section we explore the low and high temperature behavior of the entanglement entropy for the boundary subsystem $A$ in the small charge regime for the non-extremal $RN$ black holes in $AdS_4$ bulk spacetime. However for the boundary field theory dual to charged black holes the ground state is dual to the "extremal charged black hole" as mentioned earlier. So, in the small charge regime we also study the entanglement entropy of the subsystem $A$ for the boundary field theory dual to the extremal charged black holes. At low temperatures we show that it is possible to obtain a first law like relation for the boundary field theory dual to $RN$ black holes in $AdS_4$ bulk considering the extremal black hole as the ground state.

\subsection{Extremal black hole (Zero temperature)}

The gauge/gravity duality says that for boundary field theories dual to charged black holes in the $AdS$ bulk spacetime the ground state corresponds to ``extremal black holes''. Thus here we address the computation of entanglement entropy for the subsystem $A$ of the boundary field theory dual to extremal extremal black holes in $AdS_4$ bulk spacetime. As pointed out earlier through eq.(\ref{RNtemp}) that by solving $T=0$, the horizon radius for extremal black holes may be obtained as

\begin{equation}
r_{h}=\frac{\sqrt{Q}}{{3^{\frac{1}{4}}}}.\label{extrmcond}
\end{equation}

Thus in the case of extremal black holes the small charge limit also means small horizon radius $(l\frac{\sqrt{Q}}{{3^{\frac{1}{4}}}}\leq l r_h<<1)$. If we substitute the extremality condition (\ref{extrmcond}) in the expression for the lapse function given by eq.(\ref{RNlapse1}) then for the extremal black holes $f(r)$ assumes the following form

\begin{equation}
f(r)=1-\frac{4r_{h}^3}{r^3}+\frac{3r_{h}^4}{r^4}.\label{lapseExtrm}
\end{equation}

With the above form of  the lapse function the integral for the subsystem length $l$ and the extremal area ${\cal A}$ given by equations eq.(\ref{eq4}) and eq.(\ref{eq5})becomes

\begin{eqnarray}
 l&=&\frac{2}{r_{c}}\int _0^1\frac{u^2 \left(1-\frac{4{r_h}^3 }{{r_{c}}^3}u^3+\frac{3{r_h}^4}{{r_{c}}^4}u^4\right)^{-\frac{1}{2}}}{\sqrt{1-u^4}}du,\label{leqExtrm}\\
 {\cal A}&=&2Lr_{c}\int _0^1\frac{\left(1-\frac{4{r_h}^3 }{{r_{c}}^3}u^3+\frac{3{r_h}^4}{{r_{c}}^4}u^4\right)^{-\frac{1}{2}}}{u^2\sqrt{1-u^4}}du. \label{AeqExtrm}
\end{eqnarray}

As the horizon radius $(r_h)$ is small, the black hole remains deep inside the bulk and therefore far away from the extremal surface i.e $r_{h}<<r_{c}$. Thus in this limit,  we may Taylor expand the quantity $f(u)^{-1/2}$ around $\frac{r_h}{r_c}=0$ and keep the
non-vanishing terms up to $O[(\frac{r_h}{r_c})^3 u^3]$ as

\begin{equation}
f(u)^{-\frac{1}{2}}\approx1+2\frac{r_{h}^3}{r_c^3}u^3. \label{fapproxExtrm}
\end{equation}
 
Using this approximation we evaluate the integral in eq.(\ref{leqExtrm}) to be as follows 

\begin{equation}
 l \approx \frac{2}{r_{c}}\int_0^1 \bigg(\frac{u^2}{\sqrt{1-u^4}}+\frac{2 u^5 (\frac{{r_{h}}}{{r_{c}}})^3}{\sqrt{1-u^4}}\bigg)du. \label{leqExtrm1}
 \end{equation}

The relation in eq.(\ref{leqExtrm1}) can be inverted to obtained $r_c$ in terms of boundary subsystem length $l$ as follows

\begin{equation}\label{eq1}
r_{c}=\frac{\pi  {r_{h}}^3}{2 l {r_{c}}^3}+\frac{2 \sqrt{\pi } \Gamma (\frac{3}{4})}{l \Gamma (\frac{1}{4})}+O[\frac{r_h^4}{r_c^4}].
\end{equation}

Solving the above equation perturbatively in terms of $(r_{h}l)$  we obtain

\begin{equation}
 r_{c}=\frac{1}{l}\bigg[
 \frac{2 \sqrt{\pi } \Gamma (\frac{3}{4})}{ \Gamma (\frac{1}{4})}+
 \frac{l^3 {r_{h}}^3 \Gamma (\frac{1}{4})^3}{16 \sqrt{\pi } \Gamma (\frac{3}{4})^3}+O[(r_{h}l)^4]\bigg].\label{eq1new}
\end{equation}

Similarly, for obtaining an analytic expression for the extremal area we use the same approximated form of the quantity $f(u)^{-1/2}$ from eq.(\ref{fapproxExtrm}) in eq.(\ref{AeqExtrm}) to obtain following form for the extremal area

\begin{eqnarray}
{\cal A}&\approx& 2Lr_{c}\int _0^1\frac{1+2 u^3 (\frac{{r_h}}{{r_c}})^3}{u^2\sqrt{1-u^4}}du \nonumber\\
        &\approx& 2Lr_{c}\left(\int _0^1\frac{1}{u^2\sqrt{1-u^4}}du+\int _0^1\frac{2 u^3 (\frac{{r_h}}{{r_c}})^3}{u^2\sqrt{1-u^4}}du\right).\label{AeqExtrm1}
\end{eqnarray}

From the expression of ${\cal A}$ in eq.(\ref{AeqExtrm1}) it is observed that the first term is same as the pure AdS and is divergent. Therefore, we include the UV cutoff $1/r_b$ in the integral for ${\cal A}$ and add a counter term $(-2L r_b)$ in order to obtain the finite part of the extremal area as

\begin{eqnarray}
 {\cal A}^{finite}&\approx & 2Lr_{c}\int _\frac{r_{c}}{r_{b}}^1\frac{1}{u^2\sqrt{1-u^4}}du-2Lr_{b}+2Lr_{c}\int _0^1\frac{2 u (\frac{{r_h}}{{r_c}})^3}{\sqrt{1-u^4}}du\nonumber\\
&\approx & Lr_{c}\bigg[\frac{\sqrt{\pi } \Gamma (-\frac{1}{4})}{ \Gamma (\frac{1}{4})}+
 \frac{\pi  {r_{h}}^3}{{r_{c}}^3}\bigg].\label{AeqExtrm2}
\end{eqnarray}

If we substitute for $r_{c}$ from eq.(\ref{eq1new})in eq.(\ref{AeqExtrm2}) and keep terms up to $O(r_{h}^3l^3)$ then we get
the following approximated form of the finite part of extremal area

\begin{equation}
{\cal A}^{finite}=\frac{L}{l}\bigg[\frac{\pi  \Gamma (\frac{3}{4}) \Gamma (-\frac{1}{4})}{ \Gamma (\frac{1}{4}
  )^2}+\frac{l^3 {r_{h}}^3 \Gamma (\frac{1}{4})^2 (8 \Gamma (\frac{3}{4})+ \Gamma (-\frac{1}{4}))}{32 \Gamma (\frac{3}{4})^3}
  +O[(r_{h}l)^4\bigg].\label{AeqExtrm3}
\end{equation}

Since the renormalized entanglement entropy ($S_{A}^{finite}$) is related to the finite part of the extremal area as
\begin{equation}
 S_{A}^{finite}=\frac{{\cal A}^{finite}}{4G_{N}^{(3+1)}}.
\end{equation}

The explicit expression for the entanglement entropy of subsystem $A$ in a boundary theory dual to the extremal black hole in the small charge regime may be written down as follows
\begin{equation}
S_{ A}^{finite}=\frac{1}{4G_{N}^{3+1}}\frac{L}{l}\bigg[-\frac{4\pi  \Gamma (\frac{3}{4})^2}{ \Gamma (\frac{1}{4}
  )^2}+\frac{l^3 {r_{h}}^3 \Gamma (\frac{1}{4})^2 }{8 \Gamma (\frac{3}{4})^2}.
  +O[(r_{h}l)^4\bigg]\label{EEareaExtrm}
\end{equation}

For extremal black holes we also have, $r_h^3=\frac{M^{ext}}{4}$ which follows from  eq.(\ref{MQrel}) and eq.(\ref{extrmcond}). Thus replacing $r_h^3$ by $M^{ext}$ (Mass of the extremal black hole) in eq.(\ref{EEareaExtrm}) we obtain following form of renormalized entanglement entropy

\begin{eqnarray} 
 S_{ A}^{finite}&\approx& S_A^{AdS}+k M^{ext} Ll^2,~~k=\frac{1}{4G_{N}^{3+1}}\frac{\Gamma (\frac{1}{4})^2 }{32\Gamma (\frac{3}{4})^2},\label{EEareaExtrm1}
\end{eqnarray}
where, $S_A^{AdS}$ is the entanglement entropy of the subsystem $(A)$ when the bulk theory is pure AdS \cite{Fischler:2012ca}. We see that when the charge is small, the leading contribution to the entanglement entropy of the subsystem $(A)$ in the boundary field theory dual to the extremal black hole arises from the pure $AdS$ spacetime. We will see in later subsections how this sub-leading correction term in the above equation becomes important in defining the first law like relation.

\subsection{Non-extremal charged black hole (low temperature)}

We now consider the subsystem $A$ of  the boundary field theory dual to a non-extremal charged black hole in $AdS_4$ with a small charge and at a low temperature. It is observed from the extremality bound in eq.(\ref{eq2}) that when the temperature and the charge both are small, the horizon radius is small. This implies that $Q/r_{h}^2\sim 1$ and $r_h$ is small such that, $ r_{h} << r_{c} $ is satisfied. For a non-extremal black hole the form of the lapse function $f(u)$ may be given as

\begin{equation}
 f(u)=1-(\frac{r_h}{r_c})^3 u^3-\frac{Q^2}{r_h^4}\bigg((\frac{r_h}{r_c})^3 u^3-(\frac{r_h}{r_c})^4 u^4\bigg).\label{lapseRNads}
\end{equation}

We define a new parameter $\alpha=\frac{Q^2}{r_h^4}$ to substitute for black hole charge $Q$ in eq.(\ref{lapseRNads}) and then Taylor expand the quantity $f(u)^{-1/2}$ around $\frac{r_h}{r_c}=0 $ while keeping the non vanishing terms up to $O[(\frac{r_h}{r_c})^3 u^3]$ as

\begin{equation}
f(u)^{-\frac{1}{2}}\approx1+\frac{1+\alpha}{2}(\frac{r_h}{r_c})^3 u^3.\label{lapseRNads1}
\end{equation}

Using the approximated form of the lapse function given by eq.(\ref{lapseRNads1}) in the integral (\ref{eq4}) for subsystem length $l$ we obtain

\begin{equation}
 l\approx\frac{2}{r_c}\int_0^1\frac{u^2}{\sqrt{1-u^4}}\bigg(1+\frac{1+\alpha}{2}(\frac{r_h}{r_c})^3 u^3\bigg).\label{leqnonExtrm}
\end{equation}

The relation in eq.(\ref{leqnonExtrm}) can be inverted to obtained $r_c$ in terms of boundary subsystem length $l$ which is then solved perturbatively in terms of $(r_{h}l)$ to obtain the relation between $r_c$ and $l$ as follows

\begin{equation}
 r_c=\frac{1}{l}\bigg[\frac{2\sqrt{\pi } \Gamma (\frac{3}{4})}{ \Gamma (\frac{1}{4})}+\frac{(\pi  \alpha +\pi ) l^3 {r_h}^3 \Gamma (\frac{1}{4})^3}{64\pi ^{3/2} \Gamma (\frac{3}{4})^3}+O[(r_hl)^4]\bigg]. \label{eq3}
\end{equation}

Similarly, Using the approximated form of the lapse function given by eq.(\ref{lapseRNads1})in eq.(\ref{eq4}) we obtain the following form for the extremal area

\begin{eqnarray}
{\cal A}&\approx& 2Lr_c\int_0^1\frac{1}{u^2\sqrt{1-u^4}}\bigg(1+\frac{1}{2}(\frac{r_h}{r_c})^3u^3 (1+\alpha )\bigg) \nonumber\\
        &\approx& 2Lr_c\bigg[\int_0^1\frac{1}{u^2\sqrt{1-u^4}}+\int_0^1\frac{1+\alpha}{u^2\sqrt{1-u^4}}\left(\frac{r_h^3u^3}{2 r_c^3}\right)\bigg].
\label{AeqnonExtrm}
\end{eqnarray} 

The first term in the integral given in eq.(\ref{AeqnonExtrm}) for extremal area is divergent and is same as the entanglement entropy of subsystem $A$ of the boundary field theory dual to bulk $AdS_4$ space time. Therefore, regularizing the integral for ${\cal A}$ in the same way as done for the extremal black hole case in the previous subsection, we obtain the finite part of the extremal area as

\begin{equation}
 {\cal A}^{finite}\approx L \bigg[\frac{r_c \sqrt{\pi } \Gamma (-\frac{1}{4})}{2 \Gamma (\frac{1}{4})}+\frac{\pi  {r_h}^3}{4 {r_c}^2}+\frac{\pi  \alpha  {r_h}^3}{4 {r_c}^2}\bigg].\label{AeqnonExtrm1}
\end{equation}

Substituting the expression for $r_c$ from eq.(\ref{eq3}) in eq.(\ref{AeqnonExtrm1}) and keeping terms up to $O(r_h^3)$ the  finite part of the extremal area and the entanglement entropy may be written down as

\begin{eqnarray}
{\cal A}^{finite}&=&-\frac{4L\pi  \Gamma (\frac{1}{4})^2}{l \Gamma (\frac{3}{4})^2}+\frac{l^2 L {r_h}^3 \Gamma (\frac{1}{4})^2}{32\Gamma (\frac{3}{4})^2}+\frac{\alpha  l^2 L {r_h}^3 \Gamma (\frac{1}{4})^2}{32\Gamma (\frac{3}{4})^2}+O(r_h^4 l^3),\label{AeqnonExtrm2}\\
S_A^{finite} &=&\frac{1}{4G}\frac{L}{l}\bigg[-\frac{4\pi  \Gamma (\frac{1}{4})^2}{ \Gamma (\frac{3}{4})^2}+\frac{ {r_h}^3l^3 \Gamma (\frac{1}{4})^2}{32\Gamma (\frac{3}{4})^2}+\frac{\alpha   {r_h}^3l^3 \Gamma (\frac{1}{4})^2}{32\Gamma (\frac{3}{4})^2}+O(r_h^4 l^4) \bigg].\label{SeqnonExtrm}
\end{eqnarray}
 
For non-extremal $RN$ black holes in $AdS_4$ we also have a constraint relation between radius of horizon $(r_h)$, black hole charge $(Q)$ and mass $(M)$ which follows from eq.(\ref{MQrel}) as
\begin{equation}
 r_h^3(1+\alpha)=r_h^3(1+\frac{Q^2}{r_h^4})=M.\label{MQrel2}
\end{equation}

Using the constraint relation given by eq.(\ref{MQrel2}) in the expression of entanglement entropy in eq.(\ref{SeqnonExtrm}) we obtain

\begin{equation}
 S_A^{finite}\approx S_A^{AdS}+k MLl^2,~~ k=\frac{1}{4G}\frac{\Gamma (\frac{1}{4})^2}{32\Gamma (\frac{3}{4})^2}.\label{EEareanonExtrm}
\end{equation}

We see that when charge and temperature both are small, the leading contribution in the entanglement entropy for the case of non-extremal black holes comes from $AdS$ just like for the case of extremal black holes. 

\subsection{Entanglement thermodynamics of charged black holes}

From $AdS/CFT$ correspondence it may be observed that when there is a Reissner-Nordstrom black hole present in the bulk then one has to consider extremal black hole as dual to the ground state (zero temperature state) of the boundary field theory. Thus in order to obtain the ``first law of entanglement thermodynamics" we subtract eq.(\ref{EEareaExtrm1}) from eq.(\ref{EEareanonExtrm}) to obtain

\begin{equation}
\Delta S_A=\frac{1}{T_{ent} } \Delta E_A,\label{Ethermo}
\end{equation}
where,

\begin{equation}
\Delta S_{A}=S_{A}-S_{A}^{ext},
\end{equation}

\begin{equation}
\Delta E_{A}=\int_A dxdy \  T_{tt}^{Temp\neq0}-\int_ A dxdy \  T_{tt}^{Temp=0}=\frac{Ll}{8\pi G_N}(M-M^{ext}),\label{EC}
\end{equation}

\begin{equation}
 T_{ent}=\pi\frac{\Gamma (\frac{3}{4})^2}{16\Gamma (\frac{1}{4})^2} \frac{1}{l},\label{entgtemp} 
\end{equation}
here, $T_{ent}$ is known as the entanglement temperature which matches with the Schwarzschild case
and $T_{tt}$ is the time component of the stress-energy tensor of boundary field theory which can be calculated using the prescription given in\cite{Balasubramanian:1999re}. This kind of first law of entanglement thermodynamics for boundary field theories dual to charged black holes in the bulk was also derived in  \cite{Blanco:2013joa}. There the authors have considered the grand canonical ensemble with the pure $AdS$ as the ground-state. This differs from our results as we have considered the canonical ensemble for which the ground state of the boundary field theory is dual to the extremal $AdS$ black hole in the bulk. Furthermore, the first law like relation given by eq.(\ref{Ethermo}) for the entanglement entropy may be extended to include a work term due to pressure and volume as studied in \cite{Allahbakhshi:2013rda}. However, here we are only interested in studying the dependence of the quantity $\Delta S_A$ on the quantity $\Delta E_A$ as studied in \cite{Bhattacharya:2012mi}.

In \cite{Blanco:2013joa} authors also brought out the deeper aspects of the above relation by establishing its connection with the modular Hamiltonian, relative entropy and hence, to the famous Bekenstein bound. The relative entropy is defined as the stastical measure of the distance between two different states in the Hilbert space. If the density matrices corresponding to the two states are denoted by $\rho_1$ and $\rho_0$, then the relative entropy is given by
\begin{equation}
 S(\rho_1|\rho_0)= Tr(\rho_1 \log\rho_1)-Tr(\rho_1\log\rho_0).\label{rE}
\end{equation}
The relative entropy is known to be always positive $S(\rho_1|\rho_0)\geq0$ and is zero only when the two states considered are the same ($\rho_1=\rho_0$). On the otherhand, the modular Hamiltonian is defined as the logarithm of the density matrix i.e

\begin{equation}
 \rho_0 = \frac{e^{-H}}{Tr(e^{-H})},\label{modH}
\end{equation}

where, the denominator is for the purpose of normalization ($Tr(\rho_0)=1$). These two quantities, the relative entropy and the modular Hamiltonian are related by

\begin{equation}
 S(\rho_1|\rho_0) = \Delta\langle H\rangle-\Delta S,\label{re}
 \end{equation}

\begin{equation}
where,~~ \Delta S=Tr(\rho_1 \log\rho_1)-Tr(\rho_0\log\rho_0),~~ \Delta\langle H\rangle= Tr(\rho_1 H)-Tr(\rho_0H),
\end{equation}
From eq.(\ref{re}), the authors argued that the positivity of relative entropy implies the following bound on the modular Hamiltonian
\begin{equation}
 \Delta\langle H\rangle\geq \Delta S.\label{bmodh}
\end{equation}
For the case of a spherical entangling surface as considered in \cite{Blanco:2013joa} and in a handful other cases, the modular Hamiltonian is well known. The authors therefore, showed that the ''first law of entanglement thermodynamics`` essentially means that the above bound is saturated to the leading order. The reason for this is as follows. If one assumes the relative entropy to be a smooth function, then from eq.(\ref{re}) one can see that as the two states considered approach each other the relative entropy goes to zero. Therefore, one expects the bound in eq.(\ref{bmodh}) to be saturated up to the leading order, and the higher order corrections to be negative. Unfortunately, the modular Hamiltonian is not known for an infinte strip like subsystem that we have considered. Therefore, we can not comment on the connection between the bound in eq.(\ref{bmodh}) and the relation we derived in eq.(\ref{Ethermo}). We can only guess that the change in modular Hamiltonian might be of the form given in eq.(\ref{EC}), if at all the bound is saturated.

\subsection{Non-extremal charged black hole (high temperature)}

In this section we explore the high temperature behavior of the entanglement entropy when the black hole in the bulk has a small charge. From the extremality bound it may be observed that when the temperature is high and the charge is small, the horizon radius is very large ($r_hl>>1$). As a result the quantity $\frac{Q}{\sqrt{3}r_h^2}<<1$. We will call this quantity $\delta=\frac{Q}{\sqrt{3}r_h^2}$ and Taylor expand $f(u)^{-\frac{1}{2}}$ around $\delta=0$.

\begin{equation}
 f(u)^{-\frac{1}{2}}\approx \frac{1}{\sqrt{1-\frac{{r_h}^3 u^3}{{r_c}^3}}}+\frac{3}{2}(\frac{r_h}{r_c})^3\frac{\delta ^2  u^3 (1-\frac{{r_h} u}{{r_c}})}{(1-\frac{{r_h}^3 u^3}{{r_c}^3})^{3/2})}.\label{lapseRNads2}
\end{equation}

Using the approximated form of the lapse function given by eq.(\ref{lapseRNads2}) in the integrals in eq.(\ref{eq4}) and eq.(\ref{eq5}) for subsystem length $l$ and the extremal surface area ${\cal A}$ we obtain

\begin{equation}
 l=\frac{2}{r_c}\int_0^1\frac{u^2}{\sqrt{1-u^4}}\bigg(\frac{1}{\sqrt{1-\frac{{r_h}^3 u^3}{{r_c}^3}}}+\frac{3\delta^2}{2}(\frac{r_h}{r_c})^3\frac{  u^3 (1-\frac{{r_h} u}{{r_c}})}{(1-\frac{{r_h}^3 u^3}{{r_c}^3})^{3/2})}\bigg), \label{lhit}
\end{equation}

\begin{equation}
{\cal A}=2L r_c\int_0^1\frac{u^2}{\sqrt{1-u^4}}\bigg(\frac{1}{\sqrt{1-\frac{{r_h}^3 u^3}{{r_c}^3}}}+\frac{3\delta^2}{2}(\frac{r_h}{r_c})^3\frac{  u^3 (1-\frac{{r_h} u}{{r_c}})}{(1-\frac{{r_h}^3 u^3}{{r_c}^3})^{3/2})}\bigg).\label{Ahit}
\end{equation}

As observed in\cite{Hubeny:2012ry} that the extremal surface can never penetrate the horizon in a static asymptomatically AdS black hole which implies that $r_c$ is always greater than $r_h$. As the horizon radius is large for the case being studied, $r_h$ approaches very close to the extremal surface, $r_h\sim r_c$.  Thus assuming $r_c=r_h(1+\epsilon) $, the simplified form for the boundary subsystem length $(l)$ and the finite part of the extremal area from eq.(\ref{lhit}) and eq.(\ref{Ahit}) respectively can be written down as an expansion in $\epsilon$ as follows

\begin{eqnarray}
lr_{h} &=& -\frac{1}{\sqrt{3 }}\log[3\epsilon]+c_1+\delta^2 c_2+O[\epsilon] , \label{lhit1} \\
{\cal A}^{finite}&=&Llr_h^2+Lr_h(k_1+\delta^2 k_2)+Lr_h\epsilon\bigg[k_3+\delta^2 (k_4+k_5 \log[\epsilon])\bigg]+O[\epsilon^2],\label{Ahit1}
\end{eqnarray}
here the explicit expressions for the coefficients $c_1, c_2$ and $k_i$ $\{i=1,\cdots,5\}$ may be found in the Appendix-\textbf{A} where we have also discussed the steps involved in computing the integrals in eq.(\ref{lhit}) and  eq.(\ref{Ahit}) for the boundary subsystem length $(l)$ and finite part of the extremal area ${{\cal A}^{finite}}$.  Therefore following eq.(\ref{lhit1}) and eq.(\ref{Ahit1}) the renormalized entanglement entropy($S_A^{finite}=\frac{{\cal A}^{finite}}{4G}$) of non-extremal black hole at high temperature in the small charge regime may be written down as follows
\begin{eqnarray}
 S_A^{finite} &=& Ll S_{BH}  + \frac{L}{4G} r_h(k_1+\delta^2 k_2)+\frac{L}{4G}r_h\epsilon\bigg[k_3+\delta^2 (k_4+k_5 \log[\epsilon])\bigg]+O[\epsilon^2], \label{shight}\\
 \epsilon &\approx& \varepsilon_{ent}e^{-\frac{4\pi}{\sqrt3}Tl(1+\delta^2)},
\end{eqnarray}
here, the term $S_{BH}=\frac{r_h^2}{4G_N}$ corresponds to the entropy density of the planar black hole. We observe that the first term in the eq.(\ref{shight}) scales with the area of the subsystem in the 2+1 dimensional boundary CFT and increases with the temperature. Therefore we deduce that it corresponds to the thermal part. The second and subsequent  terms are proportional to the length of the boundary separating the subsystem (A) and its complement. Hence they contain the information about the entanglement at high temperatures. The $\epsilon$ corrections decay exponentially with the temperature just as they did for the Schwarzschild case, except that there is a small $\delta^2$ term due to the presence of the charge.

\section{ Large charge regime}
In this section we explore the behavior of the entanglement entropy in the large charge regime of the AdS black hole in the bulk. The extremality bound given in (\ref{eq2}) shows that when the charge is large the horizon radius is also large i.e.  ($r_hl>>1$).  So we evaluate the leading contribution to the extremal surface area by expanding the lapse function $f(u)$ around $u_0=\frac{r_c}{r_h}$, which in terms of the $r$ coordinates indicates a near horizon expansion.  A similar expansion was also used to evaluate the entanglement entropy for charged AdS black holes in \cite{Tonni:2010pv}.

\subsection{Extremal black hole}
In this context we first evaluate the entanglement entropy for the extremal black hole in the large charge regime.
In order for the near horizon expansion mentioned above to hold, it is necessary to show that the term $u-u_0$ is small, 
so that the higher order terms can be neglected. Notice that in the area integral the variable  $u$ goes from $\frac{r_c}{r_b}$ to $1$. For the large charge case $r_c\sim r_h$ so that $u_0\sim1$. Since $r_c$ and $r_b$ are both large, $u$ is close to $u_0$ through out the integral and the near-horizon expansion is valid.
Taylor expanding $f(u)$ around $u_0=\frac{r_c}{r_h}$ leads to the following form for the lapse function
\begin{equation}
 f(u)=6(\frac{r_{h}}{r_{c}})^2(u-u_0)^2+ O[(u-u_0)^3],
\end{equation}
\begin{equation}
 f(u)\approx6(1-\frac{r_{h}}{r_{c}}u)^2. \label{lapseRNads3}
\end{equation}
Using the approximated form for the lapse function given by eq.(\ref{lapseRNads3}) in the integral in eq.(\ref{eq4}) for the subsystem length $l$, we obtain
\begin{equation}\label{lrhext}
 l=\frac{2}{r_{c}}\int _0^1\frac{u^2 }{\sqrt{1-u^4}}\frac{du}{\sqrt{6}(1-\frac{r_{h}}{r_{c}}u)}.
\end{equation}

 As the horizon radius is very large the turning point is at an arbitrarily close distance to the horizon i.e $r_c=r_h(1+\epsilon)$. The integral above is then performed as an expansion in $\epsilon$. The details of the computation of the integral above are given in Appendix-\textbf{B}. The final result of this integral up to the leading order given in eq.(\ref{lext}) is as follows

\begin{equation}\label{lrhex2}
 lr_h= k_l+ \frac{\pi}{\sqrt{6\epsilon}}+O[\epsilon^{\frac{1}{2}}].
\end{equation}

The expression for constant $k_l $ in the above equation is given in Appendix-\textbf{B}. 
Using the approximated form of the lapse function given by eq.(\ref{lapseRNads3}) in the integral in eq.(\ref{eq5}) for extremal surface area ${\cal A},$ we obtain

 \begin{equation}\label{aex}
 {\cal A}=2Lr_{c}\int _0^1\frac{1}{u^2\sqrt{1-u^4}}\frac{1}{\sqrt{6}(1-\frac{r_{h}}{r_{c}}u)} du.
\end{equation}

The area integral is divergent. It is regularized and the finite part is computed as an expansion in $\epsilon$. The details of the computation are in Appendix-\textbf{B}. We state the end result which is the finite part of the extremal area given in eq.(\ref{aexfi})
\begin{equation}
 {\cal A}^{finite}= Llr_{h}^2+ L r_h (K_1+K_2\sqrt{\epsilon}+K_3\epsilon+O[\epsilon^\frac{3}{2}]).
 \end{equation}
 
The coefficients $K_1$, $K_2$ and  $K_3$ are given in Appendix-\textbf{B}. $\epsilon$ is obtained to the leading order by inverting eq.(\ref{lrhex2})
\begin{equation}\label{eq 9}
 \epsilon\approx\frac{\pi^2}{6(lr_{h}-k_{l})^2}=\frac{\sqrt{3}\pi^2}{6(l\sqrt{Q}-3^{\frac{1}{4}}k_{l})^2}.
\end{equation}

Therefore the renormalized entanglement entropy of the boundary field theory dual to the extremal charged black hole in the large charge regime is given by
\begin{equation}
 S_A^{finite}=Ll S_{BH} + \frac{Lr_h}{4G}  (K_1+K_2\sqrt{\epsilon}+K_3\epsilon+O[\epsilon^\frac{3}{2}]).
\end{equation}

The first term in the above equation scales with the area of the subsystem and is extensive. Since the black hole is at zero temperature the entire contribution comes from the charge. The $\epsilon$ corrections do not decrease exponentially with the charge in this case but in a power law form as given by eq.(\ref{eq 9}) .

 \subsection{Non-extremal black hole  }
 In this section we explore the entanglement entropy of the boundary CFT for the large charge regime of the non-extremal charged black hole. The extremality condition sets a bound on the horizon radius [$r_h>\frac{\sqrt{Q}}{3^{\frac{1}{4}}}$]. Hence
 in this regime for large charge this leads to a large horizon radius ($r_hl>>1$). The argument for the large charge regime of the extremal black hole in the previous section remains valid also for the non extremal case being considered here. Therefore we employ the same technique of the near horizon expansion for $f(u)$ around $u_0= \frac{r_c}{r_h}$ and extract the leading contribution to the area integral. The lapse function in this case is given as

\begin{equation}
 f(u)=(-3+\frac{Q^2}{r_h^4})\frac{r_h}{r_c}(u-u_0)+O[(u-u_0)^2],
\end{equation}

\begin{equation}
 f(u)\approx(3-\frac{Q^2}{r_h^4})(1-\frac{r_h}{r_c}u).\label{lapseRNads4}
\end{equation}

Note that the prefactor appearing in the above equation is the same as that which occurs in the expression for the temperature and is denoted as $\delta=(3-\frac{Q^2}{r_h^4})$. When we have $\delta\rightarrow0$ the temperature is low and when $\delta\rightarrow3$ temperature is high. Using the approximated form of the lapse function given by eq.(\ref{lapseRNads4}) in the integral in eq.(\ref{eq4}) for subsystem length $l$, we obtain

\begin{equation}
 l=\frac{2}{r_c\sqrt{\delta}}\int_0^1 \frac{u^2}{\sqrt{1-u^4}}\frac{1}{\sqrt{1-\frac{r_h}{r_c}u}}.
\end{equation}

Using the expansion given in eq.(\ref{exp1}) with $x=\frac{r_h}{r_c}u$, the integral can be evaluated 

\begin{equation}
 lr_c=\frac{1}{2\sqrt{\delta}}\sum_{n=0}^{\infty}\frac{\Gamma(n+\frac{1}{2})}
 {\Gamma(n+1)}\frac{\Gamma(\frac{n+3}{4})}
 {\Gamma(\frac{n+5}{4})}(\frac{r_h}{r_c})^n.\label{llar}
\end{equation}

The series in the above equation goes as $\sim \frac{x^n}{n}$  and therefore diverges as $r_c\rightarrow r_h$ . We isolate the divergent term. As the horizon radius is large due to the large charge regime, we set $r_c=r_h(1+\epsilon) $ and expand in $\epsilon$ to extract the leading term. We start with 
\begin{equation*}
\sqrt{\delta} lr_h=-\log[\epsilon]+k +O[\epsilon],
\end{equation*}

\begin{equation}
 \epsilon\approx\varepsilon_{ent}e^{-\sqrt{\delta}lr_{h}},
\end{equation}
where the constants $ \varepsilon_{ent}$ and $k$ are given by
\begin{equation*}
 \varepsilon_{ent}=e^{k},  \ \ \ \ k=\frac{\sqrt{\pi}\Gamma(\frac{3}{4})}{2\Gamma(\frac{5}{4})}+
 \sum_{n=1}^{\infty}\bigg(\frac{\Gamma(n+\frac{1}{2})}
 {2\Gamma(n+1)}\frac{\Gamma(\frac{n+3}{4})}
 {\Gamma(\frac{n+5}{4})}-\frac{1}{n}\bigg).
\end{equation*}
  
 Now using the approximated form of the lapse function given by eq.(\ref{lapseRNads4}) in the integral in eq.(\ref{eq5}) for the extremal surface area ${\cal A},$ we obtain
The area
\begin{equation}\label{anex}
 {\cal A}=\frac{2Lr_c}{\sqrt{\delta}}\int_0^1 \frac{1}{u^2\sqrt{1-u^4 }}\frac{1}{\sqrt{1-\frac{r_h }{r_c}u}}.
\end{equation}

The area integral above is divergent. The details of the regularization and the computation are presented in Appendix-\textbf{C}. The final result is the finite part of the extremal area is given by eq.(\ref{anextfi}) which we reproduce here
\begin{equation}
 {\cal A}^{finite}= Llr_h^2+\frac{2Lr_h}{\sqrt{\delta}}\bigg[K_1'+K_2'\epsilon+O[\epsilon^2]\bigg],
\end{equation}
where $K_1'$ and $K_2'$ are constants given in Appendix-\textbf{C}. The renormalized entanglement entropy ($ S_A^{finite}= \frac{{\cal A}^{finite}}{4G}$) for the non-extremal black hole in the large charge regime, therefore comes out to be as follows 

\begin{equation}
 S_A^{finite}= LlS_{BH}+\frac{Lr_h}{2G\sqrt{\delta}}\bigg[K_1'+K_2'\epsilon+O[\epsilon^2]\bigg],
\end{equation}

\begin{equation}
 \epsilon\approx\varepsilon_{ent}e^{-\sqrt{\delta}lr_{h}}=\varepsilon_{ent}e^{-\frac{4\pi Tl}{\sqrt{\delta}}}.
\end{equation}

The first term scales with  the area of the subsystem and is extensive. $\delta<3$ as the quantity $\frac{Q^2}{r_h^4}$ is always positive. Therefore $\epsilon$ corrections are high when the temperature is low and they decay exponentially with temperature when the temperature is high.

\section{Summary and Conclusions}

In summary we have studied the entanglement entropy of a strip like region denoted by the subsystem $A$ for the boundary conformal field theory dual to bulk charged black holes in an $AdS_4/CFT_3$ scenario. We have mainly focused on two aspects of the holographic entanglement entropy. First is the behavior of the entanglement entropy with the charge of the AdS black hole and second its temperature dependence. For this we have obtained an analytic expression for the holographic entanglement entropy using the   analytic techniques adopted in \cite{Fischler:2012ca}. In the small charge regime we have obtained the low and high temperature behavior of the holographic entanglement entropy for the boundary conformal field theory in the case of a bulk charged Reissner-Nordstrom black hole in an $AdS_4$ spacetime. As stated earlier the zero temperature ground state of the boundary field theory in this case corresponds to the extremal RN-AdS black hole in the bulk. In this context we have obtained the ``First law of entanglement thermodynamics" for the small charge and low temperature regime in the canonical ensemble. This is in contrast to the work in \cite{Blanco:2013joa} where  the first law of entanglement thermodynamics was also derived for the boundary field theory dual to charged black holes in the bulk in the grand canonical ensemble with the pure $AdS$ spacetime dual to the ground state. We have also studied the holographic entanglement entropy in the large charge regime for both non-extremal and extremal charged  black holes in the $AdS_4$ bulk spacetime. Specifically, the large charge of the black hole in the bulk forces the boundary field theory to have a high temperature in the case of non-extremal black hole whereas, in the case of extremal black hole the large charge  implies a large horizon radius . We also establish that at high temperatures the holographic entanglement entropy shows a characteristic exponential dependence on the temperature $T$ and the extremality parameter $\delta=Q/\sqrt{3}r_h^2$ both in the small and the large charge regimes. This exponential dependence of holographic entanglement entropy on temperature was also established in \cite{Fischler:2012ca} for the case of Schwarzschild black hole in the the $AdS$ bulk spacetime. Thus it is important to note that the observations made here provide a deeper understanding of entanglement and related phenomena in strongly coupled boundary conformal field theories at finite temperature and charge density.

Our work leads to extremely interesting future directions for investigations. One of the possible avenues for this is to to investigate the holographic entanglement entropy for the boundary field theories dual to rotating and charged rotating black holes in four or higher dimensions in the bulk. It would be also interesting to extend the analytic approach adopted here for the case of boundary field theories dual to charged ``Gauss-Bonnet" black holes or the R-charged black holes \cite{Dey:2014voa} in the bulk and study the temperature dependence of holographic entanglement entropy.  

\section{Acknowledgment}
The authors would like to thank Sayantani Bhattacharyya, Jyotirmoy Bhattacharya and Kaushik Ray for interesting and useful discussions. PC and VM would also like to acknowledge useful discussion with Robert Myers during {\it Strings 2015 }.
The work of Pankaj Chaturvedi is supported by the Grant No. 09/092(0846)/2012-EMR-I, from the Council of Scientific and Industrial Research (CSIR), India.
\begin{appendices}
  \section{Non-extremal charged black hole (high temperature and small charge regime)}
 In this case the integrals for boundary subsystem length and the extremal area given by eq.(\ref{lhit}) and eq.(\ref{Ahit}) respectively  may be simplified using the following identities
    
\begin{eqnarray}
 \frac{1}{\sqrt{1-x}}=\sum_{n=0}^\infty \frac{\Gamma(n+\frac{1}{2})}{\sqrt{\pi}\Gamma(n+1)} x^n, &&
  \frac{1}{{(1-x)}^{\frac{3}{2}}}=\sum_{n=0}^\infty \frac{2\Gamma(n+\frac{3}{2})}{\sqrt{\pi}\Gamma(n+1)} x^n, \label{exp1}
\end{eqnarray}

\begin{equation}
 \int_0^1 x^{\mu-1}(1-x^\lambda)^{\nu-1}=\frac{ B( \frac{ \mu }{ \lambda },\nu) }{\lambda}, \ where \ B(x,y)=\frac{\Gamma(x)\Gamma(y)}{\Gamma(x+y)},\label{ga}
\end{equation}
here eq.(\ref{ga}) corresponds to the integral representation of the beta function.

With the help of above identities the integral for the boundary subsystem length given in eq.(\ref{lhit}) may be obtained as follows

\begin{eqnarray}\label{eq8}
\frac{ lr_c}{2} &=& \sum_{n=0}^\infty \frac{\Gamma(n+\frac{1}{2})}{\sqrt{\pi}\Gamma(n+1)}(\frac{r_h}{r_c})^{3n}
 \int_0^1\frac{u^{3n+2}}{\sqrt{1-u^4}}+\delta^2 \sum_{n=0}^\infty \frac{3\Gamma(n+\frac{3}{2})}{\sqrt{\pi}\Gamma(n+1)}(\frac{r_h}{r_c})^{3n+3}
 \int_0^1\frac{u^{3n+5}(1-\frac{r_h}{r_c}u)} {\sqrt{1-u^4}}\nonumber \\
  &=& \sum_{n=0}^\infty \frac{\Gamma(n+\frac{1}{2})}{2\Gamma(n+1)}\frac{\Gamma(\frac{3n+3}{4})}{\Gamma(\frac{3n+5}{4})}(\frac{r_h}{r_c})^{3n}
 +\frac{3\delta^2}{2}\sum_{n=0}^\infty \frac{\Gamma(n+\frac{3}{2})}{\Gamma(n+1)}\frac{\Gamma(\frac{3n+6}{4})}{\Gamma(\frac{3n+8}{4})}(\frac{r_h}{r_c})^{3n+3}\nonumber \\
 &-&\frac{3\delta^2}{2}\sum_{n=0}^\infty \frac{\Gamma(n+\frac{3}{2})}{\Gamma(n+1)}\frac{\Gamma(\frac{3n+7}{4})}{\Gamma(\frac{3n+9}{4})}(\frac{r_h}{r_c})^{3n+4}.
\end{eqnarray}

In the above expression for subsystem length $l$ the first series goes as $\sim \frac{x^n}{n}$ for large n and the other two series go as $ \sim x^n $. Isolating the divergent terms, we see that the divergences of the last two series cancel each other giving us

\begin{eqnarray}
 lr_c &=& \frac{\sqrt{\pi}}{2}\frac{\Gamma(\frac{3}{4})}{\Gamma(\frac{5}{4})}+ \sum_{n=1}^\infty\bigg( \frac{\Gamma(n+\frac{1}{2})}{2\Gamma(n+1)}\frac{\Gamma(\frac{3n+3}{4})}{\Gamma(\frac{3n+5}{4})}-\frac{1}{\sqrt{3}n}\bigg)(\frac{r_h}{r_c})^{3n}\nonumber\\ 
 &+&\frac{3\delta^2}{2}\sum_{n=0}^\infty \bigg( \frac{\Gamma(n+\frac{3}{2})}{\Gamma(n+1)}\frac{\Gamma(\frac{3n+6}{4})}{\Gamma(\frac{3n+8}{4})}-\frac{2}{\sqrt{3}}\bigg)(\frac{r_h}{r_c})^{3n+3} \nonumber\\
 & -&\frac{3\delta^2}{2}\sum_{n=0}^\infty\bigg( \frac{\Gamma(n+\frac{3}{2})}{\Gamma(n+1)}\frac{\Gamma(\frac{3n+7}{4})}{\Gamma(\frac{3n+9}{4})}-\frac{2}{\sqrt{3}}\bigg)(\frac{r_h}{r_c})^{3n+4} \nonumber\\
  & +&\sqrt{3}\delta^2 \frac{(\frac{r_h}{r_c})^3}{(1+\frac{r_h}{r_c}+(\frac{r_h}{r_c})^2)}
 -\frac{1}{\sqrt{3 }}\log[1-(\frac{r_h}{r_c})^3].
\end{eqnarray}

For brevity the expression for boundary subsystem length obtained in the above equation may be written down as follows

\begin{equation}\label{lhitfi}
 lr_{h} = -\frac{1}{\sqrt{3 }}\log[3\epsilon]+c_1+\delta^2 c_2+O[\epsilon] ,
\end{equation}
here the coefficients $c_1$ and $c_2$ have following forms

\begin{eqnarray}
 c_1 &=& \frac{\sqrt{\pi}}{2}\frac{\Gamma(\frac{3}{4})}{\Gamma(\frac{5}{4})}+ \sum_{n=1}^\infty\bigg( \frac{\Gamma(n+\frac{1}{2})}{2\Gamma(n+1)}\frac{\Gamma(\frac{3n+3}{4})}{\Gamma(\frac{3n+5}{4})}-\frac{1}{\sqrt{3}n}\bigg), \nonumber \\
  c_2 &=& \frac{1}{\sqrt{3}}-\frac{3}{2}\sum_{n=0}^\infty \bigg( \frac{\Gamma(n+\frac{3}{2})}{\Gamma(n+1)}\frac{\Gamma(\frac{3n+6}{4})}{\Gamma(\frac{3n+8}{4})}-\frac{2}{\sqrt{3}}\bigg) +\frac{3}{2}\sum_{n=0}^\infty\bigg( \frac{\Gamma(n+\frac{3}{2})}{\Gamma(n+1)}\frac{\Gamma(\frac{3n+7}{4})}{\Gamma(\frac{3n+9}{4})}-\frac{2}{\sqrt{3}}\bigg) \nonumber.\\
\end{eqnarray}

From the integral given in eq.(\ref{Ahit}) for the extremal area one can see that only the first term corresponds to the Pure AdS part with divergence thus following \cite{Fischler:2012ca} we may write the finite part of the extremal area as

\begin{equation}
{\cal A}^{finite}=2L r_c\int_{\frac{r_c}{r_b}}^1\frac{u^2}{\sqrt{1-u^4}}\bigg(\frac{1}{\sqrt{1-\frac{{r_h}^3 u^3}{{r_c}^3}}}+\frac{3\delta^2}{2}(\frac{r_h}{r_c})^3\frac{  u^3 (1-\frac{{r_h} u}{{r_c}})}{(1-\frac{{r_h}^3 u^3}{{r_c}^3})^{3/2})}\bigg) -2Lr_{b}.
\end{equation}

Once again using the identities in eq.(\ref{exp1}) and eq.(\ref{ga})  the above integral reduces to following form
\begin{eqnarray}
 {\cal A}^{finite}&=& 2Lr_c\bigg[\sqrt{\pi}\frac{\Gamma(-\frac{1}{4})}{4\Gamma(\frac{1}{4})}+ \sum_{n=1}^\infty \frac{\Gamma(n+\frac{1}{2})}{\sqrt{\pi}\Gamma(n+1)}(\frac{r_h}{r_c})^{3n} \int_0^1\frac{u^{3n-2}}{\sqrt{1-u^4}}\nonumber\\
 & +&\delta^2 \sum_{n=0}^\infty \frac{3\Gamma(n+\frac{3}{2})}{\sqrt{\pi}\Gamma(n+1)}(\frac{r_h}{r_c})^{3n+3}
 \int_0^1\frac{u^{3n+1}(1-\frac{r_h}{r_c}u)} {\sqrt{1-u^4}}\bigg] \nonumber\\
  &=& \ Lr_c\bigg[\sqrt{\pi}\frac{\Gamma(-\frac{1}{4})}{2\Gamma(\frac{1}{4})}+\sum_{n=1}^\infty \frac{\Gamma(n+\frac{1}{2})}{2\Gamma(n+1)}\frac{\Gamma(\frac{3n-1}{4})}{\Gamma(\frac{3n+1}{4})}(\frac{r_h}{r_c})^{3n} \nonumber \\ 
 & +&\frac{3\delta^2}{2}\sum_{n=0}^\infty \frac{\Gamma(n+\frac{3}{2})}{\Gamma(n+1)}\frac{\Gamma(\frac{3n+2}{4})}{\Gamma(\frac{3n+4}{4})}(\frac{r_h}{r_c})^{3n+3} -\frac{3\delta^2}{2}\sum_{n=0}^\infty \frac{\Gamma(n+\frac{3}{2})}{\Gamma(n+1)}\frac{\Gamma(\frac{3n+3}{4})}{\Gamma(\frac{3n+5}{4})}(\frac{r_h}{r_c})^{3n+4}\bigg].
\end{eqnarray}

Using the identity $\Gamma(n+1)=n\Gamma(n)$ in the above equation we obtain
\begin{eqnarray}
{\cal A}^{finite} &=& \ Lr_c\bigg[\sqrt{\pi}\frac{\Gamma(-\frac{1}{4})}{2\Gamma(\frac{1}{4})}+\sum_{n=1}^\infty\bigg(1+\frac{2}{3n-1}\bigg) \frac{\Gamma(n+\frac{1}{2})}{2\Gamma(n+1)}\frac{\Gamma(\frac{3n+3}{4})}{\Gamma(\frac{3n+5}{4})}(\frac{r_h}{r_c})^{3n}\nonumber\\
&+&\frac{3\delta^2}{2}\sum_{n=0}^\infty\bigg(1+\frac{2}{3n+2}\bigg) \frac{\Gamma(n+\frac{3}{2})}{\Gamma(n+1)}\frac{\Gamma(\frac{3n+6}{4})}{\Gamma(\frac{3n+8}{4})}(\frac{r_h}{r_c})^{3n+3} \nonumber\\
 &-&\frac{3\delta^2}{2}\sum_{n=0}^\infty\bigg(1+\frac{2}{3n+3}\bigg) \frac{\Gamma(n+\frac{3}{2})}{\Gamma(n+1)}\frac{\Gamma(\frac{3n+7}{4})}{\Gamma(\frac{3n+9}{4})}(\frac{r_h}{r_c})^{3n+4}\bigg].
\end{eqnarray}

We see that we can now use eq.(\ref{eq8}) to simplify the above equation as
\begin{eqnarray}
{\cal A}^{finite} &=& \ Lr_c\bigg[\sqrt{\pi}\frac{\Gamma(-\frac{1}{4})}{\Gamma(\frac{1}{4})}+lr_c+ \sum_{n=1}^\infty\bigg(\frac{2}{3n-1}\bigg) \frac{\Gamma(n+\frac{1}{2})}{2\Gamma(n+1)}\frac{\Gamma(\frac{3n+3}{4})}{\Gamma(\frac{3n+5}{4})}(\frac{r_h}{r_c})^{3n}\nonumber\\
& +&\frac{3\delta^2}{2}\sum_{n=0}^\infty\bigg(\frac{2}{3n+2}\bigg) \frac{\Gamma(n+\frac{3}{2})}{\Gamma(n+1)}\frac{\Gamma(\frac{3n+6}{4})}{\Gamma(\frac{3n+8}{4})}(\frac{r_h}{r_c})^{3n+3} \nonumber\\
&-&\frac{3\delta^2}{2}\sum_{n=0}^\infty\bigg(\frac{2}{3n+3}\bigg) \frac{\Gamma(n+\frac{3}{2})}{\Gamma(n+1)}\frac{\Gamma(\frac{3n+7}{4})}{\Gamma(\frac{3n+9}{4})}(\frac{r_h}{r_c})^{3n+4}\bigg]. 
\end{eqnarray}

It may be observed that the first series in the above equation goes as $\sim \frac{x^n}{n^2}$ for large n, where as the last two go as  $\sim \frac{x^n}{n}$. All of these series diverge at the $O[\epsilon] $
\begin{eqnarray}
{\cal A}^{finite} &=& \ Lr_c\bigg[\sqrt{\pi}\frac{\Gamma(-\frac{1}{4})}{\Gamma(\frac{1}{4})}+lr_c+ \sum_{n=1}^\infty\bigg(\frac{1}{3n-1} \frac{\Gamma(n+\frac{1}{2})}{\Gamma(n+1)}\frac{\Gamma(\frac{3n+3}{4})}{\Gamma(\frac{3n+5}{4})}-\frac{2}{3\sqrt{3}n^2}\bigg)(\frac{r_h}{r_c})^{3n}\nonumber\\
&+&3\delta^2\sum_{n=1}^\infty\bigg(\frac{1}{3n+2} \frac{\Gamma(n+\frac{3}{2})}{\Gamma(n+1)}\frac{\Gamma(\frac{3n+6}{4})}{\Gamma(\frac{3n+8}{4})}-\frac{2}{3\sqrt{3}n}\bigg)(\frac{r_h}{r_c})^{3n+3} \nonumber\\
&-& 3\delta^2\sum_{n=1}^\infty\bigg(\frac{1}{3n+3} \frac{\Gamma(n+\frac{3}{2})}{\Gamma(n+1)}\frac{\Gamma(\frac{3n+7}{4})}{\Gamma(\frac{3n+9}{4})}-\frac{2}{3\sqrt{3}n}\bigg)(\frac{r_h}{r_c})^{3n+4}\nonumber\\
 &+& 3\delta^2\bigg((\frac{r_h}{r_c})^3\frac{(\Gamma(\frac{3}{2}))^2}{2}  -(\frac{r_h}{r_c})^4\frac{\Gamma(\frac{3}{2})\Gamma(\frac{7}{4})}{3\Gamma(\frac{9}{4})}\bigg) \nonumber\\ 
 &+&\frac{2}{3\sqrt{3}}Li_2[(\frac{r_h}{r_c})^3]-\frac{2\delta^2}{\sqrt{3}}(\frac{r_h}{r_c})^3(1-\frac{r_h}{r_c})\log[1-(\frac{r_h}{r_c})^3]\bigg].
\end{eqnarray}

After substituting $r_c=r_h(1+\epsilon)$ in the above expression for the extremal area and expanding up to $O[\epsilon]$ its explicit form may be written down as follows 
\begin{equation}\label{ahitfi}
 {\cal A}^{finite}=Llr_h^2+Lr_h(k_1+\delta^2 k_2)+Lr_h\bigg[k_3\epsilon+\delta^2 (k_4\epsilon+k_5\epsilon \log[\epsilon])+O[\epsilon^2]\bigg],
 \end{equation}
 
here the coefficients, $k_i$ $\{i=1,\cdots,5\}$ are given by following expressions
\begin{eqnarray*}
 k_1 &=& \sum_{n=1}^\infty\bigg(\frac{1}{3n-1} \frac{\Gamma(n+\frac{1}{2})}{\Gamma(n+1)}\frac{\Gamma(\frac{3n+3}{4})}{\Gamma(\frac{3n+5}{4})}-\frac{2}{3\sqrt{3}n^2}\bigg),\\
  k_2 &=& \frac{3\pi}{8}-\frac{\Gamma(\frac{3}{2})\Gamma(\frac{7}{4})}{\Gamma(\frac{9}{4})}\\
 &+&3\sum_{n=1}^\infty\bigg(\frac{1}{3n+2} \frac{\Gamma(n+\frac{3}{2})}{\Gamma(n+1)}\frac{\Gamma(\frac{3n+6}{4})}{\Gamma(\frac{3n+8}{4})}-\frac{1}{3\sqrt{3}n}\bigg),\\
 &-& 3\sum_{n=1}^\infty\bigg(\frac{2}{3n+3} \frac{\Gamma(n+\frac{3}{2})}{\Gamma(n+1)}\frac{\Gamma(\frac{3n+7}{4})}{\Gamma(\frac{3n+9}{4})}-\frac{2}{3\sqrt{3}n}\bigg)\\
 k_3&=&\frac{2\log[3]-2}{\sqrt{3}}, ~~~k_5 = \frac{1}{3},\\
 k_4 &=& -\frac{2}{\sqrt{3}}+\frac{2}{3}\log[3]-\frac{3\pi}{4}+\frac{2\Gamma(\frac{3}{2})\Gamma(\frac{7}{4})}{\Gamma(\frac{9}{4})}.
 \end{eqnarray*}

\section{Extremal black holes (Large charge regime)}
In this appendix, we compute the integral for subsystem length and area of the extremal surface given by eq.(\ref{lrhext}) and eq.(\ref{aex}) respectively. As already discussed in section 4.4, $r_{c}$ will always remain greater than $r_{h}$ and $\frac{r_{h}}{r_{c}}u<1$. We use the binomial expansion of $(1-\frac{r_{h}}{r_{c}}u)^{-1}$  in the integral given in eq.(\ref{lrhext})  to get the following expression
\begin{equation}
 \frac{l r_{c}}{2}=\frac{1}{\sqrt{6}}\sum^\infty_{n=0} (\frac{r_{h}}{r_{c}})^n\int _0^1\frac{u^{2+n} }{\sqrt{1-u^4}}du,
\end{equation}
\begin{equation}
 lr_{c}=\sum^\infty_{n=0} \sqrt{\frac{\pi }{6}}\frac{\Gamma (\frac{n+3}{4})}
 {2 \Gamma  (\frac{n+5}{4})} (\frac{{r_{h}}}{{r_{c}}})^n.\label{eq14}
\end{equation}
We see that the series goes like $\sim \frac{x^n}{\sqrt{n}}$ for large n and  diverges as $r_{c}\rightarrow r_{h}$. So we isolate the
divergent term.

\begin{equation}
 lr_{c}=\sqrt{\frac{\pi }{6}}\frac{\Gamma (\frac{3}{4})}
 {2 \Gamma  (\frac{5}{4})}+ \sqrt{\frac{\pi }{6}}\sum^\infty_{n=1}\bigg(\frac{\Gamma (\frac{n+3}{4})}
 {2 \Gamma  (\frac{n+5}{4})}-\frac{1}{\sqrt{n}}\bigg)(\frac{{r_{h}}}{{r_{c}}})^n+ 
 \sqrt{\frac{\pi }{6}}Li_{\frac{1}{2}}[\frac{r_{h}}{r_{c}}].
\end{equation}
where $Li$ is the polylog function. We use $r_{c}=r_{h}(1+\epsilon) $ and implement an expansion in $\epsilon$ and retain the leading term which leads to the following expression

\begin{equation}
 lr_{h}=\sqrt{\frac{\pi }{6}}\frac{\Gamma (\frac{3}{4})}
 {2 \Gamma  (\frac{5}{4})}+ \sqrt{\frac{\pi }{6}}\sum^\infty_{n=1}\bigg(\frac{\Gamma (\frac{n+3}{4})}
 {2 \Gamma  (\frac{n+5}{4})}-\frac{1}{\sqrt{n}}\bigg)+\sqrt{\frac{\pi }{6}}[\zeta \left(\frac{1}{2}\right)+\frac{\sqrt{\pi }}{\sqrt{\epsilon }}]+O[\epsilon^{\frac{1}{2}}].
\end{equation}

Where $\zeta(x)$ is the Riemman Zeta function. The above equation is written for brevity as follows

\begin{equation}\label{lext}
 lr_h= k_l+ \frac{\pi}{\sqrt{6\epsilon}}+O[\epsilon^{\frac{1}{2}}]
\end{equation}

The constant $ k_l $ in the above equation is given by the following expression
\begin{eqnarray}
  k_{l} &=&\sqrt{\frac{\pi}{6}}\bigg[\frac{\Gamma (\frac{3}{4})}
 {2 \Gamma  (\frac{5}{4})}+\sum^\infty_{n=1}\bigg(\frac{\Gamma (\frac{n+3}{4})}
 {2 \Gamma  (\frac{n+5}{4})}-\frac{1}{\sqrt{n}}\bigg)+\zeta (\frac{1}{2})\bigg].\nonumber
\end{eqnarray}

Having obtained the integral for subsystem length, we go on to evaluate the integral for area given in eq.(\ref{aex}). We expand$(1-\frac{r_{h}}{r_{c}}u)^{-1}$ in the integral binomially and use the identity in eq.(\ref{ga}) to obtain

\begin{equation}
 {\cal A}=2Lr_{c}\sum^\infty_{n=0} \int _0^1\frac{u^{n-2}}{\sqrt{1-u^4}}(\frac{{r_{h}}}{{r_{c}}})^n.\label{eq12}
\end{equation}

The terms corresponding to $n=0$ and $n=1$ in the above equation seem to be divergent. Both of them can be handled by introducing a cut-off $(r_b)$.  Finite part of the $n=0$ term is given by

\begin{equation}
  {\cal A}_0^{finite} =\frac{2Lr_{c}}{\sqrt{6}} \int _\frac{r_{c}}{r_{b}}^1\frac{u^{-2}}{\sqrt{1-u^4}}-\frac{2Lr_{b}}{\sqrt{6}}=2Lr_{c}\bigg[\frac{\sqrt{\frac{\pi }{6}} \Gamma (-\frac{1}{4})}{4 \Gamma (\frac{1}{4})}\bigg].\label{eq10}
\end{equation}

The term corresponding to $n=1\ \ $ in eq.(\ref{eq12}) is

\begin{eqnarray}
 {\cal A}_1&=& 2Lr_{h} \int _0^1\frac{u^{-1}}{\sqrt{1-u^4}}\\
	    &=&2Lr_{h}\sum^\infty_{k=0} \frac{\Gamma(k+\frac{1}{2})}{\sqrt{\pi}\Gamma(k+1)}\int _0^1 u^{-1+4k}.
\end{eqnarray}

We see that $k=0$ term has to be regulated.
\begin{eqnarray}
 {\cal A}_{1} &=& \frac{2Lr_{h}}{\sqrt{6}}\int_{\frac{r_{c}}{r_{b}}}^1\frac{1}{u}+\frac{2Lr_{h}}{\sqrt{6}}\sum^\infty_{k=1} \frac{\Gamma(k+\frac{1}{2})}{\sqrt{\pi}\Gamma(k+1)}\int _0^1 u^{-1+4k}\\
	&=&\frac{2Lr_{h}}{\sqrt{6}}\bigg[-\log[\frac{r_{c}}{r_{b}}]+\frac{\log[4]}{4}\bigg].
\end{eqnarray}

We know that $r_c\sim r_h $ and $r_h$ is large. Since $r_c$  and $r_b$ both are large, the quantity $-\log[\frac{r_{c}}{r_{b}}]\sim 0$. Therefore this term can be ignored.
\begin{equation}
 {\cal A}_1^{finite}\approx \frac{2Lr_{h}}{\sqrt{6}}\bigg[\frac{\log[4]}{4}\bigg].\label{eq13}
\end{equation}

Substituting the finite parts of $n=0$ and $n=1$ terms given by eq.(\ref{eq10}) and eq.(\ref{eq13}) in eq.(\ref{eq12}) we get the finite part of the area of extremal surface
\begin{equation}
 {\cal A}^{finite}=\frac{2Lr_{c}}{\sqrt{6}}\bigg[\frac{\sqrt{\pi } \Gamma (-\frac{1}{4})}{4 \Gamma (\frac{1}{4})}
+\frac{r_{h}}{r_{c}}\frac{\log[4]}{4}+\sum^\infty_{n=2} \frac{\sqrt{\pi }\Gamma (\frac{n-1}{4})}{4 \Gamma (\frac{n+1}{4})}
(\frac{{r_{h}}}{{r_{c}}})^n\bigg].
\end{equation}

The series in the above expression goes as $\sim \frac{x^n}{\sqrt{n}}$ for large n and diverges as $r_{c}\rightarrow r_{h}$.  This can be slightly re-arranged using the identity $\Gamma(n+1)=n\Gamma(n)$ to avoid divergence and get the leading contribution.
\begin{equation}
  {\cal A}^{finite}=\frac{2Lr_{c}}{\sqrt{6}}\bigg[\frac{\sqrt{\pi } \Gamma (-\frac{1}{4})}{4 \Gamma (\frac{1}{4})}
+\frac{r_{h}}{r_{c}}\frac{\log[4]}{4}+\sum^\infty_{n=2}
(1+\frac{2}{n-1})\frac{\sqrt{\pi }\Gamma (\frac{n+3}{4})}{4 \Gamma (\frac{n+5}{4})}
(\frac{{r_{h}}}{{r_{c}}})^n\bigg].
\end{equation}

We can now use eq.(\ref{eq14}) for the first term in the series of the above equation and this gives us
\begin{equation}\label{eq11}
  {\cal A}^{finite}=\frac{2Lr_{c}}{\sqrt{6}}\bigg[-2\frac{\sqrt{\pi } \Gamma (\frac{3}{4})}{ \Gamma (\frac{1}{4})}
+\frac{r_{h}}{r_{c}}\frac{\log[4]}{4}+\sqrt{6}\frac{lr_{c}}{2}
-\sqrt{\pi}\frac{1}{4\Gamma(\frac{3}{2})}+\sum^\infty_{n=2}
(\frac{1}{n-1})\frac{\sqrt{\pi }\Gamma (\frac{n+3}{4})}{2 \Gamma (\frac{n+5}{4})}
(\frac{{r_{h}}}{{r_{c}}})^n\bigg]/
\end{equation}

 The series is now convergent as it goes as $\sim \frac{x^n}{n\sqrt{n}}$ for large n. So the leading term is given by just putting $ r_{c}=r_{h} $

\begin{equation}
 {\cal A}^{finite}=Llr_{h}^2+ Lr_{h}C,
\end{equation}
where,
\begin{equation}
 C=\frac{2}{\sqrt{6}}\bigg[-2\frac{\sqrt{\pi } \Gamma (\frac{3}{4})}{ \Gamma (\frac{1}{4})}+\frac{\log[4]}{4}-\frac{1}{2}+\sum^\infty_{n=2}
(\frac{1}{n-1})\frac{\sqrt{\pi }\Gamma (\frac{n+3}{4})}{2 \Gamma (\frac{n+5}{4})}\bigg].
\end{equation}

To find the sub-leading term we put $r_{c}=r_{h}(1+\epsilon)$ and expand the finite part of the area in $\epsilon$, keeping the terms up to  $O[\epsilon] $. Even though the series in eq.(\ref{eq11}) is finite at the leading order, it is divergent at the sub-leading order in $\epsilon$. This is because the series in eq.(\ref{eq11}) is of the form $\sim \sum_n f(n)(1+\epsilon)^{-n} $ and if we expand $(1+\epsilon)^{-n}$ binomially, we see that what converges at $O[\epsilon]$ need not at $O[\epsilon^2]$. Therefore we have to check for the convergence order by order in $\epsilon$. Here we isolate the divergence in the series upto $O[\epsilon]$.

\begin{eqnarray*}
  {\cal A}^{finite}&=&\frac{2L}{\sqrt{6}}\bigg[-2\frac{\sqrt{\pi } \Gamma (\frac{3}{4})r_{c}}{ \Gamma (\frac{1}{4})}
+ r_{h}(\frac{\log[4]}{4})+\sqrt{6}\frac{lr_{c}^2}{2} \\
&-&\sqrt{\pi}\frac{r_{c}}{4\Gamma(\frac{3}{2})}+\frac{\sqrt{\pi }r_{c}}{2}\sum^\infty_{n=2}
(\frac{1}{n-1}\frac{\Gamma (\frac{n+3}{4})}{ \Gamma (\frac{n+5}{4})}-\frac{2}{n\sqrt{n}})
(\frac{{r_{h}}}{{r_{c}}})^n-\sqrt{\pi } r_{h}+\sqrt{\pi }r_{c} Li_{\frac{3}{2}}[\frac{r_{h}}{r_{c}}]\bigg].
\end{eqnarray*}

Expanding the above equation in $\epsilon$ and simplifying, we obtain

\begin{equation}\label{aexfi}
  {\cal A}^{finite}= Llr_{h}^2+ L r_h (K_1+K_2\sqrt{\epsilon}+K_3\epsilon+O[\epsilon^\frac{3}{2}]).
\end{equation}

The coefficients in the above equation, $K_i$ $(i=1,2,3)$ are given by the expressions below

\begin{eqnarray}
K_1 &=& \frac{2}{\sqrt{6}}\bigg[-2\frac{\sqrt{\pi } \Gamma (\frac{3}{4})}{ \Gamma (\frac{1}{4})}+\frac{\log[4]}{4}-\frac{1+2\sqrt{\pi}}{2}+ \sqrt{\pi } \zeta \left(\frac{3}{2}\right)+\frac{\sqrt{\pi }}{2}\sum^\infty_{n=2}
(\frac{1}{n-1}\frac{\Gamma (\frac{n+3}{4})}{ \Gamma (\frac{n+5}{4})}-\frac{2}{n\sqrt{n}})\bigg]\nonumber \\
 K_2 &=& -\frac{2\pi}{\sqrt{6}}\nonumber \\
  K_3&=&\frac{2}{\sqrt{6}}\bigg[\frac{1}{2}-\sqrt{\pi}+
\sqrt{\pi }  \zeta \left(\frac{3}{2}\right)\bigg].\nonumber
\end{eqnarray}

\section{Non-extremal black holes (Large charge regime)}

In this appendix we evaluate the extremal area integral for non-extremal black hole given in eq(\ref{anex}). Using the expansion for $\frac{1}{\sqrt{1-x}} $ given in eq.(\ref{exp1}) with $x=\frac{r_h}{r_c}u$ in the area integral, we get
  \begin{equation}
    {\cal A}=\frac{2Lr_c}{\sqrt{\delta}}\sum_{n=0}^{\infty}\frac{\Gamma(n+\frac{1}{2})}
 {\sqrt{\pi}\Gamma(n+1)}\int_0^1\frac{u^{n-2}}{\sqrt{1-u^4}}(\frac{r_h}{r_c})^n.\label{n4}
  \end{equation}
  
We see that terms corresponding to $n=0$ and $n=1$ are divergent. We isolate the divergences of these terms and regulate them with the cutoff ($r_b$). First the $n=0$ term

\begin{equation}
 {\cal A}_0=\frac{2Lr_c}{\sqrt{\delta}}\int_0^1\frac{1}{u^2\sqrt{1-u^4}}du,
\end{equation}

\begin{equation*}
 {\cal A}_0=\frac{2Lr_c}{\sqrt{\delta}}\sum_{k=0}^{\infty}\frac{\Gamma(k+\frac{1}{2})}
 {\sqrt\pi\Gamma(k+1)}\int_0^1u^{-2+4k}du.
\end{equation*}

Divergence is in the term corresponding to $k=0$ and has to be regulated with a cutoff $r_b$ .
\begin{equation}
  {\cal A}_0=\frac{2Lr_c}{\sqrt{\delta}}\int_{\frac{r_c}{r_b}}^1u^{-2}du+\frac{2Lr_c}{\sqrt{\delta}}\sum_{k=1}^{\infty}\frac{\Gamma(k+\frac{1}{2})}
 {\sqrt\pi\Gamma(k+1)}\frac{1}{4k-1}.
\end{equation}

The finite part of the above expression is given by
\begin{equation}
 {\cal A}_0^{finite}=  {\cal A}_0-\frac{2Lr_b}{\sqrt{\delta}}=\frac{2Lr_c}{\sqrt{\delta}}\bigg[-\frac{\sqrt{\pi}\Gamma(\frac{3}{4})}{\Gamma(\frac{1}{4})}\bigg].\label{a0f}
\end{equation}

Now consider the term corresponding to n=1 in eq.(\ref{n4})

\begin{eqnarray}
 {\cal A}_1 &=& \frac{Lr_h}{\sqrt{\delta}}\int_0^1\frac{u^{-1}}{\sqrt{1-u^4}} du\\
	&=&\frac{Lr_h}{\sqrt{\delta}}\bigg[\int_\frac{r_c}{r_b}^1\frac{1}{u}+\sum_{k=1}^\infty\frac{\Gamma(k+\frac{1}{2}}{ \sqrt{\pi} \Gamma(k+1)}\frac{1}{4k}\bigg]\nonumber\\
	&=&\frac{Lr_h}{\sqrt{\delta}}\bigg[\log[\frac{r_b}{r_c}]+\frac{\log[4]}{4}\bigg].\nonumber
\end{eqnarray}

Just like large charge case of extremal black hole, the quantity $\log[\frac{r_b}{r_c}]\sim0$. This is because both $r_b$ and $r_c$ are both large.  Therefore we ignore it.
\begin{equation}
 {\cal A}_1^{finite}=\frac{Lr_h}{\sqrt{\delta}}\bigg[\frac{\log[4]}{4}\bigg].\label{a1f}
\end{equation}

Substituting the finite parts of the terms corresponding to $n=0$ and $n=1$ as given by (\ref{a0f}) and eq.(\ref{a1f}) in eq.(\ref{n4}) we obtain the finite part of the area to be
\begin{eqnarray}
 {\cal A}^{finite}&=& {\cal A}_0^{finite}+ {\cal A}_1^{finite} + \frac{2Lr_c}{\sqrt{\delta}}\sum_{n=2}^{\infty}\frac{\Gamma(n+\frac{1}{2})}
 {\sqrt{\pi}\Gamma(n+1)}\int_0^1\frac{u^{n-2}}{\sqrt{1-u^4}}(\frac{r_h}{r_c})^n \\
	&=&\frac{2Lr_c}{\sqrt{\delta}}\bigg[-\frac{\sqrt{\pi}\Gamma(\frac{3}{4})}{\Gamma(\frac{1}{4})}+\frac{r_h}{2r_c}(\frac{\log[4]}{4})
 +\frac{1}{4}\sum_{n=2}^\infty\frac{\Gamma(n+\frac{1}{2})}{\Gamma(n+1)}\frac{\Gamma(\frac{n-1}{4})}{\Gamma(\frac{n+1}{4})}(\frac{r_h}{r_c})^n\bigg].
\end{eqnarray}

Using $\Gamma(n+1)=n\Gamma(n)$, we can write the above equation as follows

\begin{equation}
 {\cal A}^{finite}=\frac{2Lr_c}{\sqrt{\delta}}\bigg[-\frac{\sqrt{\pi}\Gamma(\frac{3}{4})}{\Gamma(\frac{1}{4})}+\frac{r_h}{2r_c}(\frac{\log[4]}{4})
 +\frac{1}{4}\sum_{n=2}^\infty(1+\frac{2}{n-1})\frac{\Gamma(n+\frac{1}{2})}{\Gamma(n+1)}\frac{\Gamma(\frac{n+3}{4})}{\Gamma(\frac{n+5}{4})}(\frac{r_h}{r_c})^n\bigg].
\end{equation}

We can now use eq.(\ref{llar}) for the first term of the series in the above equation and simplify to get

\begin{eqnarray}
 {\cal A}^{finite}=\frac{2Lr_c}{\sqrt{\delta}}\bigg[-\frac{\sqrt{\pi}\Gamma(\frac{3}{4})}{\Gamma(\frac{1}{4})}+\frac{r_h}{2r_c}(\frac{\log[4]}{4})+\frac{lr_c \sqrt{\delta}}{2}
-\frac{\sqrt{\pi}\Gamma(\frac{3}{4}) }{4\Gamma(\frac{5}{4})}-\frac{1}{4} \\ 
+\frac{1}{2}\sum_{n=2}^{\infty}\frac{1}{n-1}\frac{\Gamma(n+\frac{1}{2})}{\Gamma(n+1)}\frac{\Gamma(\frac{n+3}{4})}{\Gamma(\frac{n+5}{4})}(\frac{r_h}{r_c})^n\bigg].\label{aff}
 \end{eqnarray}

The series in the above equation goes as $ \sim \frac{x^n}{n^2}$ for large n and hence converges as $r_c\rightarrow r_h$. Therefore the leading term can be found just by putting $r_c=r_h$.

\begin{equation}
 {\cal A}^{finite}=Llr_h^2+\frac{Lr_h}{\sqrt{\delta}}C,
\end{equation}
where,
\begin{equation}
 C=-\frac{4\sqrt{\pi}\Gamma(\frac{3}{4})}{\Gamma(\frac{1}{4} )}+\frac{\log[4]-2}{4} +\sum_{n=2}^{\infty}\frac{1}{n-1})\frac{\Gamma(n+\frac{1}{2})}{\Gamma(n+1)}\frac{\Gamma(\frac{n+3}{4})}{\Gamma(\frac{n+5}{4})}.
\end{equation}

We saw in Appendix-\textbf{B} that, for such a series one has to check for the convergence order by order in $\epsilon$. The series in eq.(\ref{aff}) is convergent up to the leading term but diverges if we consider $O[\epsilon]$ terms.  In order to obtain the area up to $O[\epsilon]$, we isolate the divergent part in eq.(\ref{aff}) .
\begin{eqnarray}
 {\cal A}^{finite}=\frac{2L}{\sqrt{\delta}}\bigg[-\frac{2r_c\sqrt{\pi}\Gamma(\frac{3}{4})}{\Gamma(\frac{1}{4} )}+r_h\frac{\log[4]}{8}+\frac{lr_c^2 \sqrt{\delta}}{2}-\frac{r_c}{4}+
\\\frac{r_c}{2}\sum_{n=2}^\infty \bigg(\frac{1}{n-1}\frac{\Gamma(n+\frac{1}{2})}{\Gamma(n+1)}\frac{\Gamma(\frac{n+3}{4})}{\Gamma(\frac{n+5}{4})}-\frac{2}{n^2}\bigg)(\frac{r_h}{r_c})^n
-r_h+r_c \ Li_{2}[\frac{r_h}{r_c}]\bigg].
\end{eqnarray}

Now we put $r_c=r_h(1+\epsilon)$, expand in $\epsilon$ and keep the terms up to $O[\epsilon]$. The finite part of the area is given by

\begin{equation}\label{anextfi}
 {\cal A}^{finite}= Llr_h^2+\frac{2Lr_h}{\sqrt{\delta}}\bigg[K_1'+K_2'\epsilon+O[\epsilon^2]\bigg].
\end{equation}
with coefficients $K_1'$ and $K_2'$ given by the following expressions
\begin{eqnarray*}
 K_1'&=&-\frac{2\sqrt{\pi}\Gamma(\frac{3}{4})}{\Gamma(\frac{1}{4} )}+\frac{\log[4]-10}{8} +\frac{1}{2}\sum_{n=2}^\infty\bigg(\frac{1}{n-1}\frac{\Gamma(n+\frac{1}{2})}{\Gamma(n+1)}\frac{\Gamma(\frac{n+3}{4})}{\Gamma(\frac{n+5}{4})}-\frac{2}{n^2}\bigg)+\frac{\pi^2}{6}\\
    K_2'&=& \frac{\pi^2 }{6}- \frac{7}{4}.
 \end{eqnarray*}
 
\end{appendices}

\bibliographystyle{unsrt}
\bibliography{eernnewbib}
\end{document}